# Prevalence of Food and Housing Insecurity among Direct Support Professionals in New York

Jennifer Cohen, Miami University, Oxford, OH, USA and
University of the Witwatersrand, South Africa
ORCID 0000-0001-7131-8372

Yana van der Meulen Rodgers, Rutgers University, New Brunswick, NJ, USA
ORCID 0000-0001-7669-2857

**Accepted manuscript version. Cite as:** Cohen, Jennifer, and Yana van der Meulen Rodgers. "Prevalence of Food and Housing Insecurity among Direct Support Professionals in New York," Disability and Health Journal, 18 (2), April 2025, 101769. https://doi.org/10.1016/j.dhjo.2024.101769

**Abstract**

Background. Low earnings are associated with household insecurity. Direct Support Professionals (DSPs) provide support for people with intellectual and developmental disabilities, typically for wages close to state minimums, and may experience insecurity.

Objective. The purpose of the study was to evaluate the prevalence of food and housing insecurity among DSPs.

Methods. We conducted a statewide, cross-sectional survey of DSPs in New York State (2022-2023). Measures included detailed questions about food and housing insecurity. We used chi-square analyses and logistic regressions to examine relationships between insecurity and demographic characteristics as proxies for social determinants of health. A total of 4,503 DSPs responded to the survey. The analytic sample contained 2,766 respondents with complete data for all relevant variables.

Results. Overall, 62.6% experienced food and/or housing insecurity, with over half of those respondents experiencing both. Insecurity was highest among DSPs with a disability (76.2%), DSPs of color (75.7%), and those with lower income (72.4%), but over 50% of DSPs across demographic groups experienced insecurity.

Conclusions. The widespread insecurity this study demonstrates is an occupational hazard that reduces worker welfare. At the macro-level, household insecurity is a critical threat to the stability of the care and support delivery system. The human services sector is projected to grow rapidly in the future. If growth continues along low wage lines, it implies an equally rapid expansion of worker insecurity. Government action to raise pay and interventions that enhance food and housing security are needed to support workers in the care delivery system for people with intellectual and developmental disabilities.

**Introduction**

Many individuals with intellectual and developmental disabilities require personalized support to live, work, and participate in their communities.[1] Direct Support Professionals (DSPs) provide this assistance. In the U.S., there is an ongoing national shortage of DSPs and rapid turnover – over 40% in 2022.[2-4] Recent guidance stresses the importance of the DSP workforce, describing the staffing crisis as a question of human rights for those with intellectual and developmental disabilities.[5,6]

Like much of the human services sector workforce, most DSPs are women, especially women of color, and pay is low.[7,8] Low pay, in turn, is a source of food and housing insecurity, which are experienced as household conditions. Economic research typically focuses on paid work alone, but the inseparability of conditions of paid work from household conditions, and vice versa, provides a clear rationale for examining the wellbeing of workers as 'whole people.'[1]* Analyses that look beyond the workplace demonstrate that household insecurity can be an occupational hazard.[11,12]

Occupational segregation renders economic insecurity in the household a gendered and racialized phenomenon: women, people of color, those with disabilities, and people with relatively low earnings tend to experience higher levels of insecurity than men, whites, those without disabilities, and those with higher earnings.[13] This ingrained aspect of the U.S. economy warrants a social analysis that situates identities in the social context that makes those identities salient. Social forces structure an environment in which it is disadvantageous to have a nonwhite race/ethnicity or to have a disability.

Insecurity in food and housing are social determinants of health that raise the risk of chronic disease among adults, depression among adolescents, and poor learning outcomes among children.[14-19] In paid work, insecurity among workers can also degrade the quality of care provided and raise turnover intentions.[20,21] For workers, the lived experience of insecurity is a stressor.

To our knowledge, no statewide or national surveys have collected detailed information on food and housing insecurity among DSPs. Few occupational studies consider workers' household conditions or the relationships between working conditions and household conditions.[22] We aimed to fill this gap by examining the prevalence of food and housing insecurity among DSPs using a cross-sectional survey of DSPs, stratifying by gender, race/ethnicity, disability status, and income.

---

[1]* Whole person health is a common phrase in debates about integrated care for individuals with intellectual and developmental disabilities. Whole person health care promotes health over the lifespan in complex social environments.[9,10]

**Methods**

*Study design*

The sample consists of novel survey data. We surveyed individuals over the age of 18 working in New York State as DSPs, including as frontline supervisors of other DSPs. Participation was voluntary. The study was reviewed and determined to be exempt from Institutional Review Board approval by [blinded for review]. The survey was piloted with DSP volunteers in September 2022, revised accordingly, and conducted between November 2022 and May 2023. It was accessed by a QR code or an emailed link to the consent form in Qualtrics. An invitation to participate was distributed by nonprofit providers and agencies that are members of the New York Disability Advocates, the National Alliance for Direct Support Professionals, and the New York State Office for People with Intellectual and Developmental Disabilities. Informed consent was obtained from all participants. Inclusion criteria were that the participant worked as a DSP in New York State and was over age 18. There are approximately 110,000 DSPs in New York State.[23] Over 4,500 DSPs responded, but we cannot calculate a response rate because the number of eligible DSPs who received the invitation is unknown. Although the sample is non-random, the large size captures variation by numerous demographic groups and regions. Our analytic sample contains 2,766 respondents with complete data for all relevant variables.

The survey's food and housing insecurity questions are drawn from several well-established surveys, including the Census Bureau's Household Pulse Survey, the Current Population Survey's Food Supplement Survey (CPS-FSS), and the Accountable Health Communities Health-Related Social Needs Screening Tool. Appendix A contains detailed question descriptions and frequencies. Note that existing data from the CPS-FSS do not permit a reliable analysis of food insecurity for DSPs in New York because there are too few observations at this granular level.

Like the core module of the CPS-FSS, we operationalized food insecurity and housing insecurity as resulting from a financial resource constraint. The food questions ask about having enough to eat, worrying food would run out before there was money to buy more, and eating smaller meals or skipping meals due to insufficient money for food. There is no standard definition of housing insecurity. Lee et al. (2021) describe housing insecurity as a continuum of experiences related to economic hardship including inability to afford rent, frequent relocation, and housing conditions like overcrowding.[14] Our survey asked about respondents' current living situation, if they have slept outside or in a precarious setting such as a car or shelter in the past 12 months, and if they have concerns about their living situation such as mold or overcrowding.

*Statistical analysis*

In the descriptive analysis, group comparisons were assessed for statistical significance in chi-square tests. Statistical analyses were performed using Stata/MP 18.0.

In the regression analysis, we used logistic models to estimate the contributing factors to (a) food insecurity, (b) housing insecurity, and (c) both types of insecurity. The dependent variables are food insecure (0,1), housing insecure (0,1), and a combined scale (a 3-point index in which 0 means secure, 1 means either food or housing insecure, and 2 means both food and housing insecure). The logistic models, binary and ordinal, relate the odds of experiencing one or both forms of insecurity to a set of characteristics and, in the full models, to a set of two-way interaction terms representing the intersectionality between being a woman, being a person of color, having income below $60,000, and/or having a disability. The interaction terms are intended as proxies that reflect interacting social systems, like racism and ablism, not individual identities. All regression coefficients are presented in the format of an odds ratio (OR).

Individual demographic characteristics are represented by a linear variable for age and dummy variables for identifying as a woman, having at least one disability, having total household income below $60,000 per year[2*], whether the respondent lives in New York City, the presence of at least one child under age 18 in the household, and for having someone else contribute to household income. Our regression specification also includes a dummy variable for belonging to a historically marginalized racial/ethnic group (nonwhite), where white non-Hispanic is the reference group.

Appendix B, Table 1 contains sample means and cell sizes for all variables and interaction terms. All cells have more than 200 respondents; the smaller cells are evident in their relatively large standard errors. All data are self-reported, and the disability indicator is constructed from a series of seven questions about hearing, vision, mobility, bathing, concentrating, doing errands, and communicating with others.[3*]

**Results**

The majority of respondents are women; most are white, non-Hispanic; most have children living with them at home; and nearly one-third have a disability (Table 1). When we examine the sample means according to the intersection of race and gender, we see substantial gaps in household income: 71.7% of women of color have income below $60,000 compared to about 50% for the other demographic groups. About 25% of respondents have more than one job, income from which is captured in the total household income variable. One-third of white women and about one-quarter of each other demographic group report a disability.

[Table 1 here]

Of the 2,766 respondents, 62.6% experience at least one form of insecurity, with almost half of those individuals experiencing both types (Table 2). Survey data show that 75.7% of nonwhite

---

[2*] In robustness checks we also tried holding a four-year college degree as an indicator of class, but this variable had little explanatory power for reasons to be explored in future work, so it is excluded from the analysis.

[3*] The questions are: Are you deaf or do you have serious difficulty hearing? Are you blind or do you have serious difficulty seeing even when wearing glasses? Because of a physical, mental or emotional condition, do you have serious difficulty concentrating, remembering, or making decisions? Do you have serious difficulty walking or climbing stairs? Do you have difficulty dressing or bathing? Because of a physical, mental, or emotional condition, do you have difficulty doing errands alone such as visiting a doctor's office or shopping? Because of a physical, mental, or emotional condition, such as autism, do you have difficulty interacting and/or communicating with others?

DSPs and 57.1% of white DSPs experience food and/or housing insecurity. Food insecurity is most prevalent among nonwhite women (63.0%) and housing insecurity is most prevalent among nonwhite men (62.9%). Over two-thirds of DSPs with children at home have food and/or housing insecurity.

The patterns observed are as expected for disability, household income, and geography: people with disabilities experience more insecurity than those without; housing and food insecurity are highest among those with the lowest household income; and the prevalence is highest among those in NYC relative to living outside of NYC. Tellingly, even among those DSPs with the highest household income (over $100,000), 40.6% experience food and/or housing insecurity. To put this result into context, New York state has the fourth highest cost of living among all states, at $49,623 for basic necessities for a typical household.[24,25]

[Table 2 here]

Figure 1 provides a descriptive intersectional analysis of the prevalence of food and housing insecurity. Among those with a household income of $60,000 or less (accounting for more than half of the respondents), over 80% of DSPs of color and over two-thirds of white DSPs experience food insecurity, housing insecurity, or both.

[Figure 1 here]

There are large, statistically significant differences by race/ethnicity for food insecurity, housing insecurity, both types of insecurity together, and for either type (Table 3). Men are more likely to be housing insecure and women are more likely to be food insecure. White men are much less likely than other groups to experience food insecurity. Nonwhite women are much more likely to be both food and housing insecure than other demographics. The largest gaps in insecurity are among women by race/ethnicity and between women of color and white men. Gaps persist by disability status (Figure 2). Among the lowest income DSPs, over 80% of those with a disability and over two-thirds of those without a disability experience insecurity. Even among the highest income DSPs, nearly 60% of those with a disability experience food or housing insecurity or both.

[Figure 2 here]

Chi-square tests show how the association with insecurity varies among demographic groups. Women and people of color with lower income are more likely to experience food insecurity. People who have lower income and a disability are slightly more likely to experience housing insecurity than those with a disability alone. Income, disability status, and race/ethnicity have the strongest relationships with insecurity. For insecurity overall, lower income alone is the most strongly associated factor, followed by disability status and race/ethnicity. Although statistically significant, the weaker associations by most intersectional groupings suggest that the individual variables have greater explanatory power than the intersections do. We return to this result and possible explanations in the regression analysis.

[Table 3 here]

Table 4 contains regression results for food insecurity, housing insecurity, and the insecurity scale. These results confirm that being a woman substantially increases the odds of having food insecurity while it lowers the odds of having housing insecurity. However, in the interaction model for food insecurity, being a woman is no longer a statistically significant predictor, and this conclusion also applies to the insecurity scale comprising both food and housing insecurity.

Across models, being nonwhite is a meaningful and statistically significant predictor of food insecurity (OR: 2.346, SE: 0.522), housing insecurity (OR: 1.496, SE: 0.332), and the insecurity scale (OR: 2.00, SE: 0.362). The same is true of household income less than $60,000 and disability status; these coefficients remain large and statistically significant even with the inclusion of the interaction terms. Having a disability, being a racial/ethnic minority, and having relatively low household income all contribute in a meaningful way to food and housing insecurity.

Among the other variables, living with a child under the age of 18 or living in New York City makes people less secure, while being older makes people more secure. Receiving a contribution to household income also helps to reduce insecurity, but the estimate loses its precision in the food insecurity interaction model.

Disability alone is a particularly strong predictor across model specifications that include interaction terms (food insecurity OR: 3.532, SE: 0.794; housing insecurity OR: 3.869, SE: 0.905; and the insecurity scale (OR: 4.463, SE: 0.974). Holding all of the other variables and the interaction terms constant, including those with disability in them, having a disability alone raises the odds of insecurity more than any other variable.

As we anticipated from the pattern of chi-square values, most of the interaction terms in the regressions are not statistically significant, except for some of the interactions with disability. Disability interacts with being nonwhite, and with having lower household income, but the direction of each is unexpected (the ORs are below 1). A potential explanation is that nonwhite or low-income people with disabilities are receiving disability-related or other needs-based financial assistance, and this effect is partially offsetting the large adverse effect of having a disability on food and housing insecurity.

There are a handful of possible explanations for the statistical insignificance of most of the interaction terms. One is that the distributions of the variables in some interaction terms may overlap to such an extent that there is little variation for the interaction term to capture. For example, the association between having both food and housing insecurity and low income and between having both food and housing insecurity and being nonwhite, *and* between low income and nonwhite are each so strong that the interaction term adds little explanatory power. Another explanation is that the cell sizes become small for the intersections, and insecurity is so common. For example, only 18 lower-income, nonwhite DSPs with a disability do *not* experience some form of insecurity compared to 134 DSPs who do. Alternative specifications using measures of acute insecurity, such as going hungry and sleeping outside, do reveal some of the expected associations for the interaction terms (see Appendix A for definitions).

[Table 4 here]

**Limitations**

Because DSPs with the longest total working hours – those working overtime; those with second, third, and even fourth jobs; and those with intensive caregiving responsibilities – are less likely to complete a time-consuming survey, our results may be lower bound estimates of the prevalence of food and housing insecurity among DSPs. Women of color in our sample (22.7%; n=629) are underrepresented compared to the state's workforce means for home health aides (72.9%) and personal care aides (49.0%),[4*] but the representation of women of color located in NYC (48.6%; n=412) is more closely aligned with workforce demographics from other surveys.[26] In the National Core Indicators report (2022), 54.3% of New York State DSP respondents are people of color; the authors note that statewide results for New York are skewed toward New York City.[27] In our sample, 30.0% of respondents are people of color, but 62.9% of DSPs in New York City are people of color. To enhance the depth of information about these issues, future surveys could be supplemented with follow-up interviews with DSPs who have multiple jobs and intensive caregiving responsibilities.

The analysis uses the gender identity question ("women (including transgender woman), men (including transgender man), and non-binary") for the gender comparisons. However, the analysis is limited to those who identified as women or men due to the small number of non-binary people among respondents. Similarly, race/ethnicity is divided into white/nonwhite to be inclusive of all respondents with complete data; cell sizes for a more refined breakdown are too small for this analysis.[28]

The disability questions included in our survey are those recommended by the Department of Health and Human Services (HHS); the same used in the American Community Survey.[29,30] They ask people to self-report if they have difficulty with basic and some instrumental activities of daily living such as seeing, hearing, walking, cognition, self-care, and communication.[30,31] The questions are the benchmark against which scholars have evaluated the Washington Group Short-Set (WG-SS).[32] Unlike the popular WG-SS, which has four response categories of severity, the HHS-recommended questions are dichotomous, with yes or no responses. Some scholars have critiqued the WG-SS questions and have characterized them as "a minimum standard measure of disability," and there is compelling comparative evidence demonstrating that the WG-SS underestimates the prevalence of disability.[32,33] In future studies, multiple definitions with variable cut points and dimensions of disability should be carefully considered.

**Discussion**

Analyzing survey data from 2,766 DSPs in New York State, we find that the majority of respondents across demographic groups experience food and/or housing insecurity. Insecurity is

---

[4*] These tabulations for New York state workforce means are based on the authors' analysis of micro-data from the 2018-2022 American Community Survey. Direct Support Professionals still do not have a single occupational code, so our comparison uses the two most closely aligned occupation categories.

a statewide problem across gender, race, income, and disability status in New York. The data further show large gaps by race/ethnicity and disability. Over two-thirds of DSPs with a disability and income of $60,000 or above experience insecurity while 82% with income below $60,000 experience insecurity. For DSPs without a disability the figures are 43% and 68%. Similarly, over 63% of DSPs of color and 46% of white DSPs with higher income experience insecurity. For those with income below $60,000, 82% of DSPs of color and 67% of white DSPs experience insecurity. In other words, DSPs with a disability and DSPs of color must earn over $60,000 in order to have the same likelihood of insecurity as DSPs without a disability and white DSPs who earn less than $60,000.

Our estimates of food insecurity for DSPs are considerably higher than the official estimate of food insecurity for all workers in New York State, which measured 24.9% in 2021 (12.8% of households in 2022).[34-36] Comparing results using the most similar single survey question across the two survey instruments, albeit with a 3-month rather than 12-month time frame in the DSP survey, food insecurity among DSP respondents was 44.1% − almost twice the level of workers at large. As one respondent noted, "I am worried about how to make ends meet and try not to spend extra money. I live paycheck to paycheck and get stressed out about food prices and feeding my family." Another wrote, "The cost of living has gone up, gas, heating, electric and food prices have gone up. But our pay is the same. [H]ow do [they] expect us to keep up when we cannot support ourselves, always worrying if the next paycheck will be able to cover the bills and have food in the house for your kids to eat."

Men (53.2%) are more likely than women (45.4%) to be housing insecure. Nationally, about 20% of households experienced housing insecurity between late April 2020 and December 2020.[37] While these estimates are not directly comparable, housing insecurity is much higher among DSPs than for people nationally. A respondent described their housing situation, "I live in an old RV [year-round] with no running water in winter. If I had to pay $1000 a month on my horrible wages, I'd be eating cat food."

Regression analysis indicates that low income, being of nonwhite, having a disability, living in NYC, and having a child in one's household all increase the odds of being insecure, while being older and having another person contribute to household income reduce the odds.

The widespread insecurity this study demonstrates is related to the low pay and low social status of the occupation: human services work is female-dominated and much of it is considered unskilled.[7,8,11,12,38] Occupational segregation has persistently crowded women and people of color, the majority of the U.S. population, into a subset of sectors and jobs, pushing wages down in those fields.[13] The human services sector is a case study in these dynamics.[39-41] People with disabilities are also more likely than those without to work in healthcare support operations and personal care and service occupations.[42] Long Covid may play a role in the high prevalence of disability among clusters of care workers, with consequences for the people for whom they provide care.[43] Devaluation is evident in the low compensation accorded to workers who provide care and support.

In national data from the Bureau of Labor Statistics, only a small number of occupations pay less than home health and personal care aides, which had a median hourly wage in 2023 of $16.12.

Childcare workers were paid $14.60 per hour on average; less than animal care and service workers (e.g., dog groomers) at $15.31 per hour.[44] Pay for direct care is low relative to almost any other job.[5*] Pay is a condition of work that, for many people, dictates household conditions, including insecurity. When pay is too low to allow a minimally-decent standard of living, people are forced to work overtime, get a second paid position, or to leave a job or field entirely.

Food and housing insecurity may be experienced privately by DSPs in their households, but it is an experience shared by the majority of respondents. Stressors that link working conditions (i.e., pay) to household conditions (i.e., insecurity), are often considered 'personal challenges.'[45] However, if some threshold level of insecurity is met by many workers in a single occupation, seemingly personal problems become an occupational hazard. As an occupational study, this finding indicates that the occupation itself is unsustainable in its current state. Because of the low pay and related understaffing, many DSPs are forced to work overtime to make ends meet, leading to an overworked burnt-out workforce, lower quality support, and persistent precarity throughout the sector.

People with intellectual and developmental disabilities and their families depend on DSPs for many kinds of support, ranging from life management work to life-or-death demands like assistance with eating. Yet unlike the market-based wages for many paid care workers, DSP pay is effectively set by Medicaid reimbursement rates in fee-for-service and managed care models. Hence their pay is a policy decision. The implication is that state and federal governments have chosen to institutionalize economic insecurity among DSPs and their households. The overrepresentation of women, especially women of color in this occupation, and of people with disabilities means that by imposing economic insecurity on DSPs, state and federal governments are also reinforcing gender, race, and disability inequities as well as the social devaluation of people with intellectual and developmental disabilities.

For policymakers, these data paint a troubling picture of the experiences of DSPs outside of paid work, which are a function of their earnings. The data also suggest that policymakers are complicit in reproducing inequities. Every effort must be made to raise DSP pay in the interest of worker well-being and the stability of the direct support delivery system. A second tack is advocacy around legislation requiring a Standard Occupational Classification (SOC) for DSPs. An SOC will permit DSP-specific analyses that are not currently possible but could help improve conditions for DSPs.

Describing this crisis in the direct support delivery system as a 'workforce crisis' belies the extraordinary problems in the direct care sector. In direct care, a 'workforce crisis' is a sectoral crisis because the workforce is the support delivery mechanism. We take care not to present a narrow, instrumentalist perspective of DSPs in this study, but it bears noting here: without a workforce, there is nothing to deliver. The sector fundamentally relies on the workforce and its well-being for the delivery of services.

---

[5*] The pattern holds in New York State, where the hourly average is $18.41 for home health and personal care aides, $18.26 for childcare workers, and $18.36 for animal care and service workers. It is $26.12 for animal trainers and $23.70 for preschool teachers.[37]

Practically, this centrality of the workforce means that worker well-being should be a key focus for employers and that well-being encompasses both paid work and life outside of paid work. Where latitude for raising pay is constrained, employers should find alternative ways of expressing the value of DSPs. This proposition is old news to many agencies that have struggled with recognition in the past. Ongoing research considers the conceptualization and operationalization of recognition.

Researchers have invested considerable personal resources in this area of study, which remains understudied and underfunded.[46] There are myriad challenges about which to conduct research and a desire from employers for data-informed analyses and interventions. We should acknowledge, however, that interventions at the micro level cannot mitigate the structural issues in the sector and tend to frame individuals or groups of DSPs as 'the problems.'

Because DSP workers are so central to the direct care sector and because their wellbeing is understudied, there are plenty of important areas for future research. One potential direction is to examine the consequences of the poverty-level wages that DSPs earn, particularly the need to work overtime or multiple jobs in order to make ends meet. Given that DSPs need additional pay but may not desire additional working hours, the extra work from overtime or another job may be more coercive than opportune. Another possible research direction is an investigation of the surprisingly high rates of food and housing insecurity among those DSPs with over $100,000 in household income. Also warranted is a closer look at our argument that economic insecurity among DSPs has been institutionalized through pay via Medicaid reimbursement. What are the channels through which DSPs effectively subsidize the state in the existing Medicaid reimbursement model? Finally, a follow-up survey would be ideal, especially one that gets an adequate response from people working for the state, who earn considerably more per hour than private-sector DSPs. One could then compare levels of insecurity and disability as well as the impact of pay on insecurity in a matched sample of respondents.

**Conclusion**

The results have consequences for public health. In particular, the food and housing insecurity among DSPs documented in this study perpetuate already large racial disparities in health outcomes. The low pay assigned to DSPs penalizes workers (and their households) who enjoy their work but are pushed out of it because of inadequate pay. Low morale and high turnover among DSPs, in turn, also penalize the recipients of the care they provide. At the macro scale, the disability services sector as a whole will be increasingly fragile if states continue to marginalize workers through this inequitable pay structure. Moving forward, the human services sector is projected to grow more rapidly than other occupations as the population ages and as disabilities become more prevalent. If growth continues along these low wage lines, it implies an equally rapid expansion of food and housing insecurity.

References

1. Mir HA, Liu Q, Rosca O, Blakeslee E. Factors That Influence the Tenure of Direct Support Professionals in New York State Provider Agencies. *Intellect Dev Disabil*. 2024;62(1):14-26. https://doi.org/10.1352/1934-9556-62.1.14

2. National Core Indicators Intellectual and Developmental Disabilities. (2023). National Core Indicators Intellectual and Developmental Disabilities State of the Workforce in 2022 Survey Report. https://idd.nationalcoreindicators.org/wp-content/uploads/2024/02/ACCESSIBLE_2022NCI-IDDStateoftheWorkforceReport.pdf

3. Tyo MB, Desroches ML. Content analysis of qualitative interviews for user-centered design of a prototype mobile health app for direct support professionals' resilience. *Disabil Health J*. 2024;17(2):101544. https://doi.org/10.1016/j.dhjo.2023.101544

4. Friedman C. The COVID-19 pandemic and quality of life outcomes of people with intellectual and developmental disabilities. *Disabil Health J*. 2021;14(4):101117. https://doi.org/10.1016/j.dhjo.2021.101117

5. America's Direct Support Workforce Crisis: Effects on People with Intellectual Disabilities, Families, Communities and the U.S. Economy. Report to the President, President's Committee for People with Intellectual Disabilities, United States Department of Health and Human Services. 2017. https://acl.gov/sites/default/files/programs/2018-02/2017%20PCPID%20Full%20Report_0.PDF

6. Leser KA, Pirie PL, Ferketich AK, Havercamp SM, Wewers ME. Dietary and physical activity behaviors of adults with developmental disabilities and their direct support professional providers. *Disabil Health J*. 2017;10(4):532-541. https://doi.org/10.1016/j.dhjo.2017.01.006

7. England P, Budig M, Folbre N. Wages of Virtue: The Relative Pay of Care Work. Social Problems. 2002;49(4):455-473. https://doi.org/10.1525/sp.2002.49.4.455

8. Baron SL, Tsui EK, Quinn MM. Work as a Root Cause of Home Health Workers' Poor Health. *Am J Public Health*. 2022;112(1):9-11. https://doi.org/10.2105%2FAJPH.2021.306582

9. National Center for Complementary and Integrative Health. Whole Person Health: What You Need to Know. 2021. https://www.nccih.nih.gov/health/whole-person-health-what-you-need-to-know

10. Thomas H, Mitchell G, Rich J, Best M. Definition of whole person care in general practice in the English language literature: a systematic review. *BMJ Open*. 2018;8(12):e023758. https://doi.org/10.1136%2Fbmjopen-2018-023758

11. Cohen J, Venter WDF. The integration of occupational- and household-based chronic stress among South African women employed as public hospital nurses. *PLoS One*. 2020;15(5):e0231693. https://doi.org/10.1371%2Fjournal.pone.0231693

12. Cohen, J. Precarity of Subsistence: Social Reproduction Among South African Nurses. *Fem Econ*. 2022;29(1):236–265. https://doi.org/10.1080/13545701.2022.2123950

13. Bergmann BR. Occupational Segregation, Wages and Profits When Employers Discriminate by Race or Sex. *East Econ J*. 1974;1(2):103–110. http://www.jstor.org/stable/40315472

14. Lee CY; Zhao X; Reesor-Oyer L; Cepni AB; Hernandez DC. Bidirectional Relationship Between Food Insecurity and Housing Instability, *J Acad Nutr Diet*. 2021;121(1):84-91. https://doi.org/10.1016/j.jand.2020.08.081

15. Walker RJ, Garacci E, Dawson AZ, Williams JS, Ozieh M, Egede LE. Trends in Food Insecurity in the United States from 2011-2017: Disparities by Age, Sex, Race/Ethnicity, and Income. *Popul Health Manag*. 2021;24(4):496-501. https://doi.org/10.1089%2Fpop.2020.0123

16. DeLuca S, Rosen E. Housing insecurity among the poor today. *Annu Rev Sociol*. 2022;48(1):343-371. https://doi.org/10.1146/annurev-soc-090921-040646

17. Pruitt SL, Leonard T, Xuan L, Amory R, Higashi RT, Nguyen OK, et al. Who Is Food Insecure? Implications for Targeted Recruitment and Outreach, National Health and Nutrition Examination Survey, 2005–2010. *Prev Chronic Dis*. 2016;13:160103. http://dx.doi.org/10.5888/pcd13.160103

18. Coley RL, Leventhal T, Lynch AD, Kull M. (2013). Relations between housing characteristics and the well-being of low-income children and adolescents. *Dev Psychol*. 2013;49(9):1775–1789. https://doi.org/10.1037%2Fa0031033

19. McIntyre L, Wu X, Kwok C, Patten SB. The pervasive effect of youth self-report of hunger on depression over 6 years of follow up. *Soc Psychiatry Psychiatr Epidemiol*. 2017;52(5):537-547. https://doi.org/10.1007/s00127-017-1361-5

20. Allen L, Williamson S, Berry E, Alderwick H. The cost of caring: poverty and deprivation among residential care workers in the UK. *Health Foundation*. 2022. https://www.health.org.uk/publications/long-reads/the-cost-of-caring

21. Srinivasan M, Cen X, Farrar B, Pooler JA, Fish T. Food Insecurity Among Health Care Workers in The US: Study examines food insecurity among health care workers in the United States. *Health Aff*. 2021:40(9):1449-1456. https://doi.org/10.1377/hlthaff.2021.00450

22. Baker J, Kruse L, Bridges A, Galindo N. DSP Survey Report. Relias/ANCOR. Accessed 2024 Aug 10. https://www.ancor.org/wp-content/uploads/2023/05/2023-Relias-DSP-Survey-Report.pdf

23. NY State Office for People with Developmental Disabilities. Accessed 2024 Aug 10. https://opwdd.ny.gov/providers/direct-support-professionals

24. Rothstein, R. Examining The Cost Of Living By State In 2024. Forbes. Accessed 2024 Sep 25. https://www.forbes.com/advisor/mortgages/cost-of-living-by-state/

25. Living Wage Calculation for New York. Accessed 2024 Sep 25. https://livingwage.mit.edu/states/36

26. Pettingell SL, Houseworth J, Tichá R, Kramme JED, Hewitt AS. Incentives, Wages, and Retention Among Direct Support Professionals: National Core Indicators Staff Stability Survey. *Intellect Dev Disabil*. 2022;60(2):113-127. https://doi.org/10.1352%2F1934-9556-60.2.113

27. National Core Indicators. (2022). National Core Indicators Intellectual and Developmental Disabilities 2020 Staff Stability Survey Report. https://www.nationalcoreindicators.org/resources/staff-stability-survey/

28. Jenkins DG, Quintana-Ascencio PF. A solution to minimum sample size for regressions. *PLoS One*. 2020 Feb 21;15(2):e0229345. https://doi.org/10.1371%2Fjournal.pone.0229345

29. U.S. Department of Health and Human Services Implementation Guidance on Data Collection Standards for Race, Ethnicity, Sex, Primary Language, and Disability Status. 2011. https://aspe.hhs.gov/sites/default/files/documents/8c4525c504acc3b1bac16586119ce729/dhhs-implementation-guidance-data-collection-standards.pdf Accessed 2024 Sep 25.

30. U.S. Centers for Disease Control and Prevention. Disability and Health - Disability Data. Disabilities in Public Health Data. https://www.cdc.gov/ncbddd/disabilityandhealth/datasets.html Accessed 2024 Sep 25.

31. Edemekong PF, Bomgaars DL, Sukumaran S, et al. Activities of Daily Living. [Updated 2023 Jun 26]. In: StatPearls [Internet]. Treasure Island (FL): StatPearls Publishing; 2024 Jan. Accessed 2024 Sep 25. https://www.ncbi.nlm.nih.gov/books/NBK470404/

32. Landes, SD, Swenor, BK, Vaitsiakhovich, N. Counting disability in the National Health Interview Survey and its consequence: Comparing the American Community Survey to the Washington Group disability measures. *Disabil Health J*. 2024;17(2):101553 https://doi.org/10.1016/j.dhjo.2023.101553

33. Brandt DE, Ho P, Chan L, Rasch EK. Conceptualizing disability in US national surveys: application of the World Health Organization's (WHO) International Classification of Functioning, Disability, and Health (ICF) framework. *Qual Life Res.* 2014; 23:2663–2671

34. Healthy People 2030. Current Population Survey Food Security Supplement (CPS-FSS). Accessed 2024 Aug 10. https://health.gov/healthypeople/objectives-and-data/data-sources-and-methods/data-sources/current-population-survey-food-security-supplement-cps-fss

35. New York State Department of Health. Self-Reported Food Insecurity Among New York State Adults by County, Behavioral Risk Factor Surveillance System. 2021. Release date: 12/29/2023. Accessed 2024 Aug 10. https://www.health.ny.gov/statistics/prevention/injury_prevention/information_for_action/docs/2023-12_ifa_report.pdf

36. Cobbs EF, McCarthy JB, A. Havusha A, Sandman D. Still Hungry: Food Insufficiency in New York State 2020-2023. New York Health Foundation. 2024. Accessed 2024 Aug 10. https://nyhealthfoundation.org/resource/food-insufficiency-in-new-york-state-2020-2023.

37. Cai JY, Fremstad S, Kalkat S. Housing Insecurity by Race and Place During the Pandemic. Center for Economic and Policy Research. 2021. Accessed 2024 Aug 10. https://cepr.net/report/housing-insecurity-by-race-and-place-during-the-pandemic/

38. Hallee Y, Parent-Lamarche A, Delattre M. Is job evaluation compatible with care work? *J Ind Relat*. 2024;66(3):331-357. https://doi.org/10.1177/00221856241254141

39. Duffy M, Albelda R, Hammonds C. Counting Care Work: The Empirical and Policy Applications of Care Theory. *Social Problems*. 2013;60(2):145-167. https://doi.org/10.1525/sp.2013.60.2.145

40. Folbre N, Gautham L, Smith K. Gender Inequality, Bargaining, and Pay in Care Services in the United States. ILR Review. 2023;76(1):86-111. https://doi.org/10.1177/00197939221091157

41. Small SF, Rodgers YV, Perry T. Immigrant women and the covid-19 pandemic: an intersectional analysis of frontline occupational crowding in the United States. *Forum for Social Economics* 2023: 1-26. https://doi.org/10.1080/07360932.2023.2170442

42. Persons with a Disability: Labor Force Characteristics – 2023. Accessed 2024 Aug 11. https://www.bls.gov/news.release/pdf/disabl.pdf

43. Cohen J, van der Meulen Rodgers Y. An intersectional analysis of long COVID prevalence. *Int J Equity Health.* 2023;22:261. https://doi.org/10.1186/s12939-023-02072-5

44. Occupational Outlook Handbook. U.S. Bureau of Labor Statistics. [Home Health and Personal Care Aides](#), [Childcare Workers](#), [Animal Care and Service Workers](#). State data under 'State & Area Data.' Accessed 2024 Aug 13. https://www.bls.gov/ooh/

45. Hansel TC, Saltzman LY, Melton PA. Work Environment and Health Care Workforce Well-Being: Mental Health and Burnout in Medically Underserved Communities Prone to Disaster. *Am J Public Health*. 2024;114(S2):156-161. https://doi.org/10.2105/ajph.2023.307478

46. Pettingell SL, Bershadsky J, Anderson LL, Hewitt A, Reagan J, Zhang A. The Direct Support Workforce: An Examination of Direct Support Professionals and Frontline Supervisors During COVID-19. *Intellect Dev Disabil*. 2022;61(3):197-210.

**Figure 1.** Food insecurity, housing insecurity, and any insecurity by household income, race/ethnicity, and gender

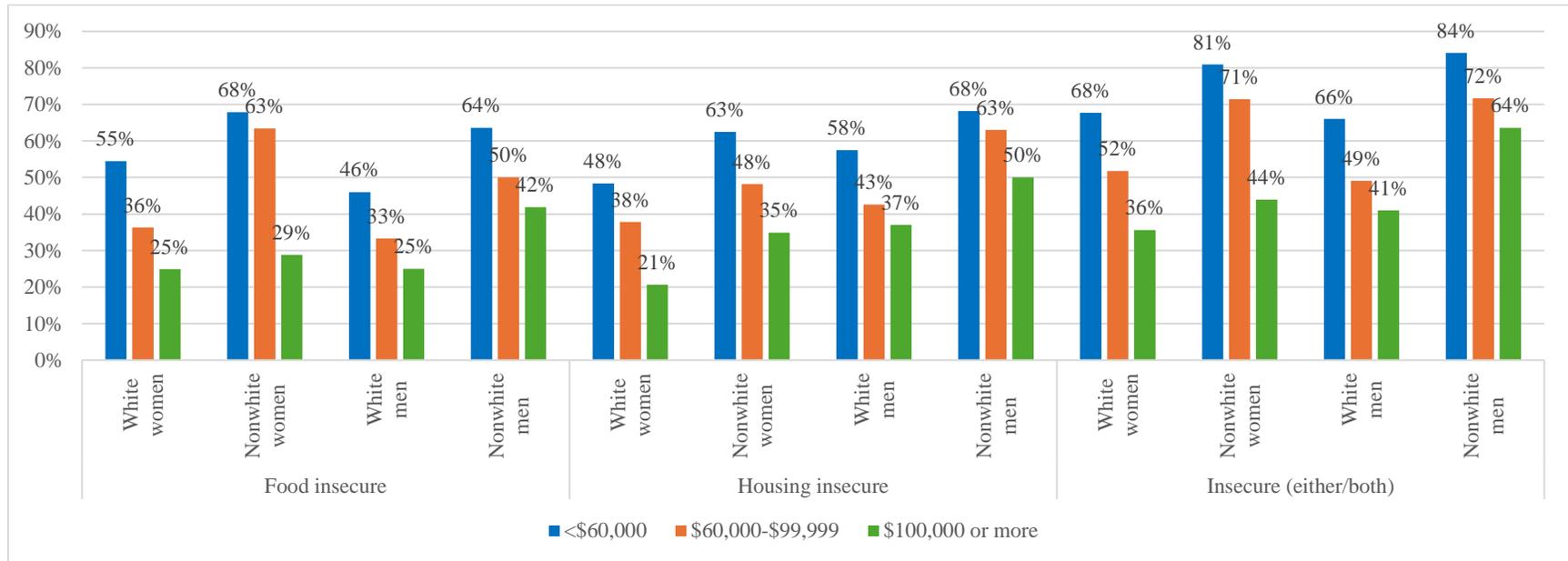

**Figure 2.** Food insecurity, housing insecurity, and any insecurity by household income and disability status

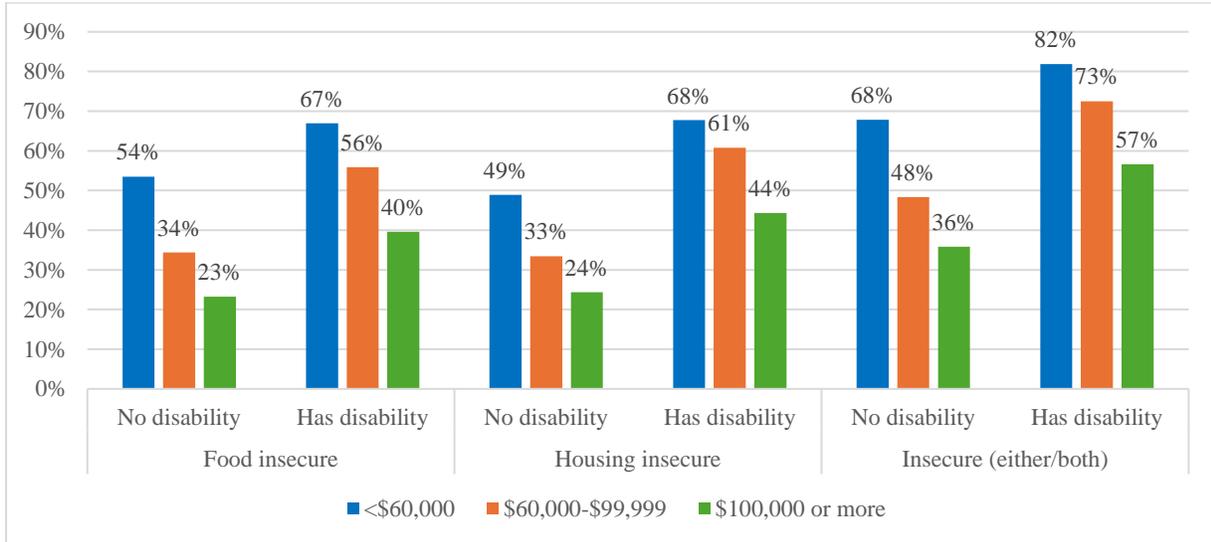

**Table 1.** Sample characteristics: DSP respondents in NYS, in percent unless indicated otherwise (N=2,766)

|  | *Total sample (mean)* | *White women* | *Nonwhite women* | *White men* | *Nonwhite men* |
|---|---|---|---|---|---|
| Gender is woman | 78.1 | | | | |
| Race is white, non-Hispanic | 70.1 | | | | |
| Race and gender interaction | | | | | |
|   White women | 55.4 | | | | |
|   Nonwhite women | 22.7 | | | | |
|   White men | 14.8 | | | | |
|   Nonwhite men | 7.1 | | | | |
| Age (in years) | 42.0 | 42.5 | 42.0 | 40.8 | 40.3 |
| Household income | | | | | |
|   <$60,000 | 56.6 | 52.7 | 71.7 | 49.0 | 54.3 |
|   $60,000-$99,000 | 26.6 | 30.7 | 17.8 | 26.5 | 23.4 |
|   >$100,000 | 16.7 | 16.5 | 10.5 | 24.5 | 22.3 |
| Location | | | | | |
|   New York City | 30.6 | 14.3 | 65.5 | 23.3 | 61.4 |
|   Outside of NYC | 69.4 | 85.7 | 34.5 | 76.7 | 38.6 |
| Has a disability | 30.1 | 33.0 | 26.2 | 27.0 | 26.9 |
| Has children at home | 67.9 | 65.3 | 72.7 | 69.6 | 69.0 |
| Receives contribution to household income | 60.3 | 62.2 | 48.6 | 71.1 | 59.9 |

Source: Authors' calculations from survey data.

**Table 2.** Prevalence of food and housing insecurity by demographic, in percent (N=2,766)

|  | *Food insecure* | *Housing insecure* | *Food and housing insecure* | *Total food and/or housing insecure* |
|---|---|---|---|---|
| All | 48.2 | 47.1 | 32.6 | 62.6 |
| Women | 49.5 | 45.4 | 32.2 | 62.7 |
| Men | 43.3 | 53.2 | 34.1 | 62.3 |
| White | 42.6 | 42.2 | 27.8 | 57.1 |
| Nonwhite | 61.1 | 58.5 | 43.9 | 75.7 |
| White women | 44.0 | 40.5 | 27.0 | 57.5 |
| White men | 37.5 | 48.5 | 30.5 | 55.4 |
| Nonwhite women | 63.0 | 57.1 | 44.7 | 75.4 |
| Nonwhite men | 55.3 | 62.9 | 41.6 | 76.7 |
| No disability | 42.8 | 40.2 | 26.2 | 56.7 |
| Disability | 60.5 | 62.9 | 47.3 | 76.2 |
| Household income |  |  |  |  |
| <$60,000 | 57.9 | 55.0 | 44.4 | 72.4 |
| $60,000-$99,000 | 40.8 | 41.7 | 26.9 | 55.6 |
| >$100,000 | 27.0 | 28.9 | 15.2 | 40.6 |
| New York City | 56.9 | 57.6 | 41.8 | 72.7 |
| Outside of NYC | 44.3 | 42.4 | 28.6 | 58.2 |
| Children in household | 52.6 | 50.7 | 35.9 | 67.3 |
| Receives contribution to household income | 43.1 | 41.8 | 28.4 | 56.5 |

Source: Authors' calculations from survey data.

**Table 3.** Chi-square tests of food and housing insecurity by demographic group (N=2,766)

|  | Food insecure | Housing insecure | Food and housing insecure | Total food and/or housing insecure |
|---|---|---|---|---|
| Nonwhite/white | 79.50*** | 61.46*** | 68.83*** | 100.67*** |
| Women/men | 7.30*** | 11.76*** | 0.81 | 0.17 |
| White women/everyone else | 23.82*** | 58.89*** | 48.06*** | 56.02*** |
| Nonwhite women/everyone else | 71.44*** | 32.70*** | 53.81*** | 71.78*** |
| White men/everyone else | 21.77*** | 0.41 | 0.98 | 6.04** |
| Nonwhite men/everyone else | 4.37** | 21.45*** | 7.89*** | 16.37*** |
| White women/white men | 5.56** | 8.44*** | 1.88 | 0.03 |
| Nonwhite women/white men | 64.38*** | 7.27*** | 20.88*** | 39.29*** |
| Nonwhite men/white men | 17.20*** | 11.09*** | 7.35** | 19.53*** |
| Nonwhite women/white women[6*] | 64.14*** | 49.22*** | 63.27*** | 81.04*** |
| Nonwhite women/nonwhite men | 3.67* | 2.13 | 0.56 | 0.13 |
| Disability/no disability | 73.43*** | 120.77*** | 117.35*** | 136.05*** |
| Income <$60,000 | 135.99*** | 90.63*** | 101.00*** | 161.37*** |
| Woman*Disability | 53.45*** | 63.45*** | 66.26*** | 83.24*** |
| Woman*Income<$60k | 113.98*** | 37.80*** | 62.20*** | 101.90*** |
| Disability*Nonwhite | 36.67*** | 65.40*** | 68.79*** | 70.81*** |
| Disability*Income<$60k | 87.84*** | 106.46*** | 115.29*** | 138.05*** |
| Income<$60k*Nonwhite | 99.69*** | 76.83*** | 86.44*** | 126.03*** |
| Woman*Disability*Income<$60k | 68.99*** | 68.16*** | 78.11*** | 97.84*** |
| Woman*Disability*Nonwhite | 28.06*** | 51.77*** | 53.72*** | 55.24*** |
| Woman*Nonwhite*Income<$60k | 83.71*** | 51.67*** | 69.36*** | 95.50*** |
| Nonwhite*Disability*Income<$60k | 39.85*** | 62.91*** | 68.44*** | 71.89*** |
| Woman*Nonwhite*Income<$60k*Disability | 28.58*** | 47.60*** | 52.63*** | 53.04*** |

The notation *** is p<0.01, ** is p<0.05, * is p<.10. The insecurity scale in column 4 uses the Kruskal-Wallis test.
Source: Authors' calculations from survey data.
[6*] The gap among women by race/ethnicity is large and significant. We report values for white women/nonwhite women and nonwhite women/nonwhite men, with the caveat that those statistics effectively "bake in" gender inequity in the former and racial inequity in the latter. For examining inequality and inequity the reference group

should be the group that is best off – the group that does not confront structural and institutional discrimination by gender and/or race, hence the other comparisons are with white men.

**Table 4** Logistic regression results for food and housing insecurity and the security scale (Odds ratios; Standard errors in parentheses) (Complete results in Appendix B)

| | *Food insecurity Simple model* | *Food insecurity Interaction model* | *Housing insecurity Simple model* | *Housing insecurity Interaction model* | *Insecurity scale Simple model* | *Insecurity scale Interaction model* |
|---|---|---|---|---|---|---|
| Woman | 1.330*** | 1.304 | .679*** | .640*** | 0.955 | 0.932 |
| | (0.133) | (0.229) | (0.068) | (0.106) | (0.087) | (0.145) |
| Nonwhite | 1.722*** | 2.346*** | 1.449*** | 1.496* | 1.652*** | 2.000*** |
| | (0.178) | (0.522) | (0.148) | (0.332) | (0.152) | (0.362) |
| Household income <$60k | 2.123*** | 2.018*** | 1.766*** | 1.638** | 2.087*** | 1.974*** |
| | (0.189) | (0.405) | (0.157) | (0.316) | (0.169) | (0.362) |
| Has disability | 2.040*** | 3.532*** | 2.695*** | 3.869*** | 2.585*** | 4.463*** |
| | (0.185) | (0.794) | (0.244) | (0.905) | (0.212) | (0.974) |
| Age | 0.970*** | 0.969*** | 0.981*** | 0.981*** | 0.972*** | 0.972*** |
| | (0.003) | (0.003) | (0.003) | (0.003) | (0.003) | (0.003) |
| Lives in NYC | 1.401*** | 1.411*** | 1.664*** | 1.674*** | 1.585*** | 1.589*** |
| | (0.142) | (0.144) | (0.169) | (0.172) | (0.145) | (0.146) |
| Child <18 years old living in household | 1.146*** | 1.463*** | 1.342*** | 1.343*** | 1.474*** | 1.477*** |
| | (0.133) | (0.144) | (0.123) | (0.123) | (0.120) | (0.121) |
| Receives financial contribution to household | 0.851* | 0.864 | 0.747*** | 0.752*** | 0.787*** | 0.796*** |
| | (0.079) | (0.081) | (0.069) | (0.070) | (0.065) | (0.066) |
| Two-way interaction terms | | | | | | |
| Woman*Nonwhite | | 0.921 | | 0.967 | | 0.941 |
| | | (0.197) | | (0.214) | | (0.181) |
| Woman*Disability | | 0.720 | | 0.815 | | 0.713 |
| | | (0.157) | | (0.189) | | (0.148) |
| Woman*Income<$60k | | 1.288 | | 1.240 | | 1.273 |
| | | (0.261) | | (0.250) | | (0.236) |
| Disability*Nonwhite | | 0.611** | | 0.917 | | 0.733* |
| | | (0.122) | | (0.199) | | (0.134) |
| Disability*Income<$60k | | 0.759 | | 0.743 | | 0.721** |
| | | (0.136) | | (0.136) | | (0.119) |
| Income<$60k*Nonwhite | | 0.812 | | 1.003 | | 0.888 |
|                   | (0.153) |         | (0.192) |         |         | (0.153) |
| Constant & Cut 1  | .919    | 0.830   | 1.080   | 1.064   | -0.715  | -0.647  |
|                   | (0.196) | (0.206) | (0.234) | (.256)  | (0.196) | (0.224) |
| Cut 2             |         |         |         |         | 0.741   | 0.815   |
|                   |         |         |         |         | (0.196) | (0.224) |
| Sample size       | 2,766   | 2,766   | 2,766   | 2,766   | 2,766   | 2,766   |
| F Statistic       | 39.35***| 23.41***| 37.73***| 21.79***| 57.28***| 33.47***|

Source: Authors' analysis of survey data.
Standard errors in parentheses. The notation *** is $p<0.01$, ** is $p<0.05$, * is $p<.10$. The insecurity scale models use ordinal logit.